\documentclass[prd,superscriptaddress,nofootinbib,notitlepage]{revtex4-1}
\usepackage{epsfig,amsmath,amssymb,slashed}
\usepackage{graphicx}
\usepackage{color}
\usepackage{siunitx}
\usepackage{color}
\usepackage{bm}
\usepackage{subfig}

\newcommand{\di}{{\rm d}}

\newcommand{\e}{{\rm e}}

\newcommand{\be}{\begin{equation}}
\newcommand{\ee}{\end{equation}}                                                                               
\def\bea{\begin{eqnarray}}
\def\eea{\end{eqnarray}}                                                                               
\begin{document}

\title{Exergy of an open continuous medium} 

\author{F. Becattini}
\affiliation{Universit\`a di Firenze and INFN Sezione di Firenze, Florence, Italy}

\begin{abstract}
Exergy is a very important thermodynamic quantity in several fields such as economy, 
engineering, ecology and yet it has attracted little attention in pure physics.
One of the main problems of the currently used definition of exergy is its dependence
on an arbitrarily chosen reference state, which is the thermodynamic state of a
reservoir the system is supposedly in contact with. In this paper, starting from a 
very general definition of exergy, a formula is derived for the exergy balance of 
a general open continuous medium without any reference to an external environment.
A formula is also derived for the most suitable thermodynamic parameters of the Earth
atmosphere when seen as an external environment in the usual exergy applications. 
\end{abstract}

\maketitle

\section{Introduction}
\label{intro}

Exergy, a term coined by Z. Rant \cite{rant}, is defined in words as "the maximum 
useful work possible during a process that brings the system into equilibrium 
with a heat reservoir, reaching maximum entropy" \cite{perot}. Indeed, the definition 
of such a quantity can be found, without receving any special naming, in the renowned 
book of statistical physics by L. Landau \cite{landau}, specifically 
in sections "Maximum work" and "Maximum work done by a body in an external medium" 
where a commonly known expression of exergy (even though not the most general, as
we will see) is quoted:
\be\label{exergy0}
 B = U + p_R V - T_R S
\ee
where $U$, $V$ and $S$ are the internal energy, volume and entropy of the system
while $p_R$ and $T_R$ are the constant pressure and temperature of the 
environment, a large thermal reservoir the system is supposedly in contact with.

Exergy, along with the expression \eqref{exergy0}, is a well known quantity in applied 
thermodynamics, with remarkable applications in various fields such as ecology, 
economics, engineering and others \cite{szargut, dincer, lizarraga}. In a world 
where sustainability is becoming the crucial problem for the future of mankind, 
exergy apparently makes it possible to quantify it \cite{jorgensen,stanek}. 
Notwithstanding, in spite of its paramount importance, exergy has attracted little 
attention in pure physics, specifically in theoretical statistical mechanics and 
thermodynamics. 
Indeed, while energy is always conserved in all physical processes, exergy does 
not, just like entropy. However, unlike entropy, exergy always {\em decreases} in 
an isolated system. This is a definitely more attractive feature with respect to
entropy for it just matches the characteristic of what we could properly call a 
resource: something that can be used for some purpose, but bound to exhaust if 
not replaced. Altogether, exergy is the concept of resource \cite{grubb}
that is actually meant in economics, politics and in the common language as well 
when energy is referred to. However, unlike energy and entropy, exergy is not
a function of the equilibrium thermodynamic state of a system. Nor is it an 
intensive and additive physical quantity, with one remarkable exception, that
of small systems in contact with an infinite reservoir, see Section~\ref{continres}. 
In fact, exergy depends on both the the system and the surrounding environment
in a non-trivial fashion, which perhaps makes it a less attractive quantity for 
physicists.

Most of the quantitative applications of exergy to date are based on the equation
\eqref{exergy0} and extensions to include the chemical part. As it has been 
mentioned, the formula \eqref{exergy0} requires the existence of a reservoir, with 
an invariably constant and uniform thermostatic intensive quantities: temperature, 
pressure, chemical potentials. However, when dealing with a continuous medium which 
is not in contact with a heat reservoir, this is a serious limitation. For instance, 
in classical thermodynamics it is well known that a fluid with temperature gradients 
within, and yet, not in contact with a reservoir, is potentially able to perform useful work 
because it has not achieved full thermodynamic equilibrium. One should be able,
in principle, to calculate the amount of this useful work. Another equally important
question is how to calculate the intensive parameters of what can be regarded as
the reservoir in most applications, i.e. the Earth atmosphere or a portion thereof.
What can be taken as reference values of the atmosphere $T_R,p_R$, given that its 
thermodynamic state is not constant neither uniform? 

To answer these questions, a more general definition of exergy is needed, one
that yields the equation \eqref{exergy0} as a special case. Indeed, a general
definition of exergy as the difference between total energy and the equilibrium 
energy for the actual value of entropy, appears in the aforementioned Landau's book 
\cite{landau} and it has been rediscovered more recently \cite{grubb} 
(see Section~\ref{isolated}). In this work, we extend this definition (see 
Section~\ref{contin}, also reported in the Conclusions) to an open 
continuous medium, i.e. not necessarily isolated and we show that the special
case of a system in contact with a reservoir is recovered. Besides, we will be
able to provide a formula to calculate the reference thermodynamic intensive 
quantities for a system which is large enough to play the role of the environment.

The paper is organized as follows: in Section~\ref{isolated} we review the fundamental
definition of exergy of an isolated system; in Section~\ref{maxwork} we illustrate
the relation between maximum work and exergy considering two finite sources at 
different temperatures; in Section~\ref{reservoir} we derive the well known expressions
of exergy of system and environment based on the general definition of exergy of
an isolated system; in Section~\ref{contin} we extend the definition of exergy 
to an open continuous medium and we derive the exergy rate equation for a multi-
component fluid; finally, in Section~\ref{atmosph} we derive a formula to estimate
the reference temperature and pressure of the Earth atmosphere.

\subsection*{Notation}

In this paper we use the natural units, with $\hbar = 1$ and the Boltzmann constant
$K_B=1$, that is temperature has the dimension of an energy.

\section{Exergy of an isolated system}
\label{isolated}

\begin{figure}[htb]
 \includegraphics[scale=0.6]{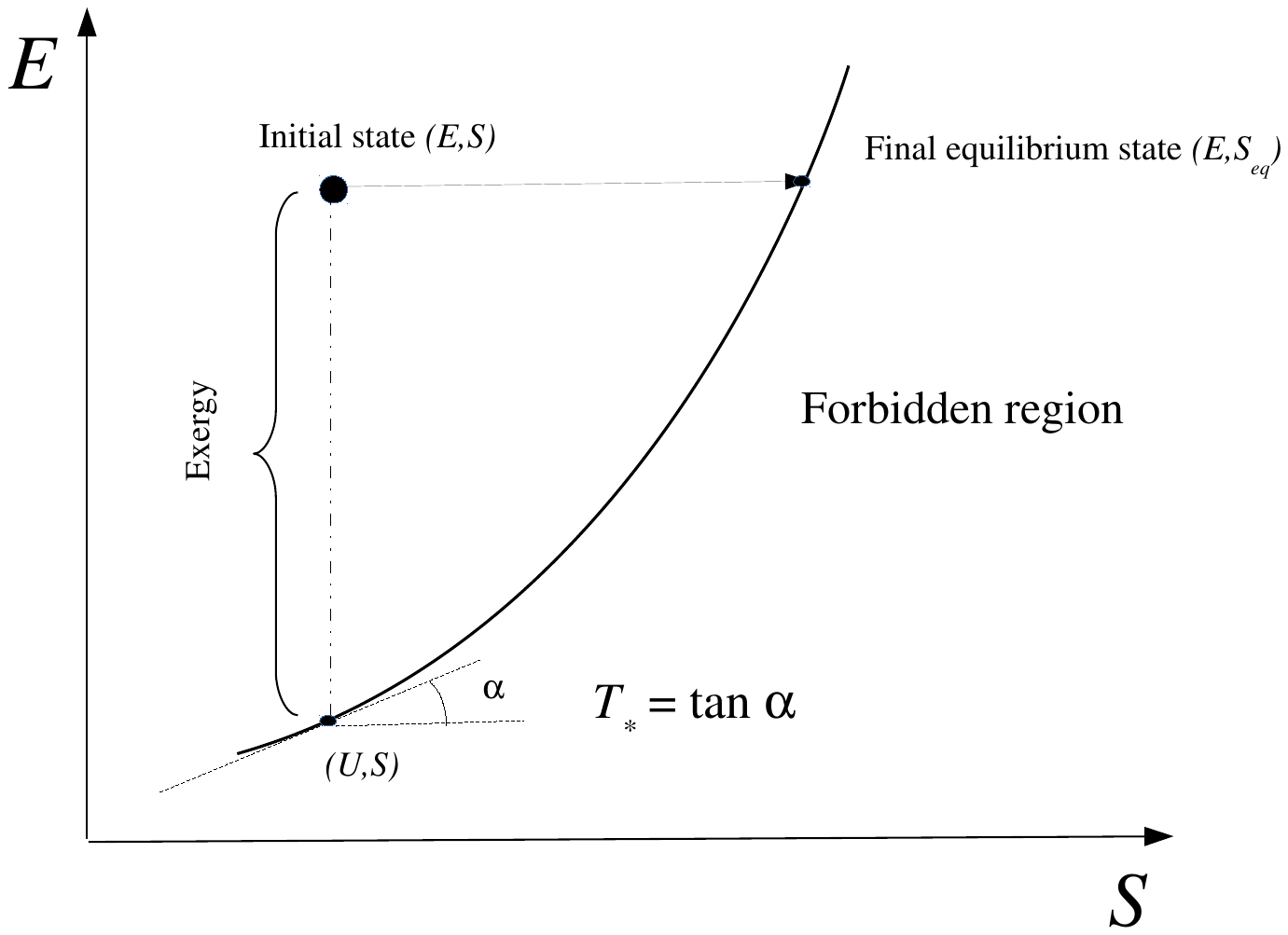}
        \caption{Energy-entropy diagram of an isolated system with fixed volume $V$.
The allowed states lie in the region above the equilibrium curve $U(S)$. The exergy
is the difference between the actual energy and the energy $U(S)$ at constant 
entropy. The irreversible evolution drives the system from the initial state $(E,S)$ 
to the equilibrium point $(E,S_{\rm eq})$, entailing a decrease of exergy until
it vanishes.}
        \label{figura0}
\end{figure}

Consider an isolated system with fixed volume $V$. Let $p_i(t)$ be the probability that
the system is in a state $i$ at the time $t$. The entropy:
$$
  S(t) =  - \sum_i p_i(t) \log p_i(t)
$$
is a function of $t$ and the second law of thermodynamics states that, after some
time, it will get to its maximum value $S_{\rm eq} = \log \Omega$ where $\Omega$ is 
the total number of microstates. It is common wisdom that the increase of entropy 
entails a degradation of the ``useful" energy. This is just the exergy, which can 
be defined, for an isolated body \cite{landau}, as the difference between the total, 
constant, energy $E_0$ and the energy that the same body would have when at equilibrium 
with total entropy $S(t)$ and volume $V$:
\be\label{exeriso}
  B(t) = E_0 - U(S(t),V)
\ee
It should be emphasized that $U(S(t),V)$ is not the {\em actual} total internal 
energy of the system, which may even coincide with $E_0$; this difference will
become clear in Section~\ref{maxwork}. The equation \eqref{exeriso} has been obtained 
starting from an equivalent definition of exergy in ref.~\cite{grubb}, 
for the case of $E_0$ only comprising internal energies (see Section~\ref{maxwork}).

It is clear that exergy decreases in time because of the second law:
\be\label{exergrate}
 \frac{\di B}{\di t} = - \frac{\partial U}{\partial S}\Big|_V \frac{\di S}{\di t}
= - T_*(t) \frac{\di S}{\di t} \le 0
\ee
where $T_*(t)$ is the temperature the system would have if thermalized at the total
entropy $S(t)$. It is important to stress that $T_*(t)$ is not the physical temperature
of the isolated system, which in principle does not even exists. In figure \ref{figura0}
$T_*(t)$ is the slope of the equilibrium function $E(S(t))$ in the point $(U(t),S)$,
and becomes a physical temperature only in the equilibrium point. For a locally 
equilibrated system $T_*(t)$ is not easy to determine, as we will see later on in
this work. However, note that the system might not even be locally equilibrated
(like a non-locally thermalized kinetic system) and yet the definition \eqref{exeriso} 
and the equation \eqref{exergrate} would hold, because the entropy argument in 
\eqref{exeriso} could be calculated anyway. 

The above expressions can be extended to an isolated system with more conserved
quantities other than volume, like momentum ${\bf P}$, electric charge $Q$ etc. 
In the definition, one just needs to include these constants in the list of the
arguments:
$$
  B(t) = E_0 - U(S(t),V,{\bf P},Q,\ldots)
$$

A very important case is the multi-component system, comprising several chemical 
species $X_k$ \footnote{This term includes either molecules or atoms or nuclei or 
subnuclear particles} which can give rise to chemical reactions. In this case, one 
should take into account that the given initial abundances $N_k(t)$ may not be at 
chemical equilibrium, so that the internal energy at fixed entropy value $S(t)$ 
should be estimated at a chemical equilibrium state, with abundances differing from 
the initial ones. A system of $M$ species subjected to $L$ reactions, features 
$M-L$ independent constants $Q_i$, which are linear functions of the $N_k$'s:
\be\label{qlinear}
 Q_i = \sum_{k=1}^M C_{ik} N_k   \qquad \qquad i=1,\ldots,M-L
\ee
Note that the $C_{ik}$'s are also constant, depending on the stoichiometric coefficients 
of the reactions, but the $N_k$'s can vary in time and yet return the same $Q_i$'s 
because the linear system of equations is not invertible; indeed, there are $M$ $N_k$'s 
but only $M-L$ $Q_i$'s.
The corresponding exergy expression involves the internal energy as a function 
of those $M-L$ independent chemical constants:
\be\label{exergy1}
  B(t) = E_0 - U(S(t),V,Q_1,\ldots,Q_{M-L})
\ee
Even though the abundances are not independent variables, as we have just discussed, 
it is sometimes useful to maintain them as redundant thermodynamic variables to
describe the internal energy:
\be\label{exergy2}
  B(t) = E_0 - U(S(t),V,{\overline N}_1(t),\ldots,{\overline N}_M(t))
\ee
It is important to stress that the ${\overline N}_k(t)$ are not the actual abundances
at the time $t$, but the {\em equilibrium abundances}, that is those the system 
would have if all chemical reactions were at equilibrium with a given total 
entropy $S(t)$. This specification is necessary because the internal energy is 
an equilibrium thermodynamic function, hence its arguments must be equilibrium
quantities as well. Note that the equilibrium abundances ${\overline N}_k(t)$ depend 
on time just because $S(t)$ does. For a single chemical reaction 
$\sum_k \nu_k X_k = 0$, $\nu_k$ being 
the stoichiometric coefficients, positive for the products and negative for the 
reactants, the chemical equilibrium condition reads:
$$
\sum_k \frac{\partial U}{\partial {\overline N}_k} \Big|_{S(t),V} \nu_k = 
\sum_k \mu_{*k}(t) \nu_k = 0
$$
where the $\mu_{*k}$ are the chemical potentials. As the internal energy can be
seen as a function of the chemical constants $Q_i$ or the equilibrium abundances
${\overline N}_k$, two kinds of chemical potentials can be defined. Nevertheless
a relation can be derived between the chemical potentials $\mu_{*iV}$ related to 
the $Q_i$'s and those $\mu_{k*V}$, related to the $\overline N_k$'s:
\be\label{chempotrel}
  \frac{\partial U}{\partial \overline N_k}\Big|_{S(t),V} = \mu_{*kV} = 
  \sum_{i=1}^{M-L} \frac{\partial U}{\partial \overline Q_i}\Big|_{S(t),V}
  \frac{\partial Q_i}{\partial \overline N_k} = \sum_{i=1}^{M-L} C_{ik} \mu_{*iV}
\ee
which can also be interpreted by saying that the $C_{ik}$ is the ``charge" of 
type $i$ carried by the species $k$.

It should be stressed that the exergy is {\em not} a function of the equilibrium
thermodynamic state. However, according to the definitions \eqref{exeriso} and its
extension \eqref{exergy2} in the microcanonical ensemble its value does not 
depend on the specific process leading to equilibrium. For most applications in
engineering, the goal is in fact the reduction of the exergy destruction rate 
\eqref{exergrate}, hence dissipation, to a minimum, so as to maintain the resource 
for a longer time.

\section{Exergy of two systems}
\label{maxwork}

One can learn much about the nature of the exergy from the simplest case of two 
adjacent systems. Suppose, for further simplicity, that they are thermalized 
subsystems of equal size and composition at temperature $T_1$ and $T_2$ with 
$T_1 < T_2$ which are separated by a non-conducting wall (see figure~\ref{figura1}). 
The system {\em as a whole} is obviously not at global thermodynamic equilibrium. 
According to the definition  we can write the total exergy at the initial time 
$t=0$ as:
$$
 B(0) = E_0 - U_{\rm th}(S_1+S_2)
$$
where the other arguments can be omitted, assuming that the volume and number of
particles do not change in time. It is important to stress that the thermalized
energy $U_{\rm th}(S_1+S_2)$ is the total energy of the system $1+2$ as though 
it was completely thermalized with entropy $S=S_1+S_2$ and it is not equal to 
$U_1 + U_2$ (see further below). As the two subsystems have equal size and
composition, in a thermalized state they have the same entropy, so we can write:
\be\label{average}
 U_{\rm th}(S_1+S_2) = U_1\left( \frac{S_1+S_2}{2} \right) + 
 U_2\left( \frac{S_1+S_2}{2} \right) 
\ee
Now, since the sum of the internal energies must be conserved in time, we can also
write:
$$
  E_0 = U_1(S_1) + U_2(S_2)
$$
so that:
\be\label{exer2}
 B(0) = U_1(S_1) + U_2(S_2) - U_1\left( \frac{S_1+S_2}{2} \right) - 
 U_2\left( \frac{S_1+S_2}{2} \right) 
\ee
This equation can be extended to $N$ bodies in thermal contact \cite{grubb}.
The equation~\eqref{exer2} demonstrates that the definition \eqref{exeriso} is not 
{\em extensive}, namely that it does not allow to express the total exergy as the 
integral over some region of a density, unlike energy and entropy. If this was 
the case, the total exergy in \eqref{exer2} would be the sum of the exergies of 
the two regions in figure~\ref{figura1}, defined according to the \eqref{exeriso}.
Since the energies in the two regions are completely thermalized, we have:
$$
  B_i = E_i - U_{\rm th}(S_i) = 0  \qquad \qquad i=1,2
$$
and yet $B(0)$ in eq.~\eqref{exer2} is not vanishing, so:
$$ 
  B(0) \ne B_1 + B_2
$$
Indeed, while $E_0 = E_1 + E_2$, in the equation \eqref{exer2} we have that:
$$
  U_{\rm th}(S_1) +  U_{\rm th}(S_2) \ne U_{\rm th} (S_1+S_2)
$$
with equality applying only if $U$ was proportional to $S$, which is generally
not the case. More generally, it can be shown that if $V_1$ and $V_2$ are two adjacent 
non-overlapping regions, the following general inequality holds:
\be\label{inequa}
  B_{V_1 \cup V_2}(t) \ge B_{V_1}(t) + B_{V_2}(t)
\ee
provided that the total energy of $V_1$ and $V_2$ can be obtained by summing the
individual energies. Since $E_{V_1 \cup V_2} = E_{V_1}+E_{V_2}$, the \eqref{inequa}
becomes:
\be\label{stability}
  U_{\rm th}(S_1+S_2,V_1+V_2+,Q_1+Q_2) \le U_{\rm th}(S_1,V_1,Q_1) + 
  U_{\rm th}(S_2,V_2,Q_2)
\ee
which is just the thermodynamic stability condition, i.e. the convexity of the
internal energy with respect to extensive quantities \cite{landau}. In the equation
\eqref{stability} $Q$ stands for the set of additive constants related to chemistry
as discussed in the Section~\ref{intro}. 
The inequality \eqref{inequa} expresses the common wisedom that exergy is 
{\em not} a function of a single system, but it substantially depends on both the 
system and its surroundings.

Altogether, this analysis shows that exergy is not an extensive quantity. While 
there is an energy and an entropy density, an exergy density cannot be defined,
with the notable exception of sytems in contact with a reservoir, see Section~\ref{continres}. 
The lack of extensivity could have been realized already from the exergy destruction 
rate in equation \eqref{exergrate}: while the entropy is extensive, the temperature 
$T_*(t)$ is non-local, in that it does not belong to some finite region, but it 
is the temperature that the {\em whole} system would have if it was thermalized. 

\subsection{Exergy as maximum work}

\begin{figure}
       \includegraphics[scale=0.4]{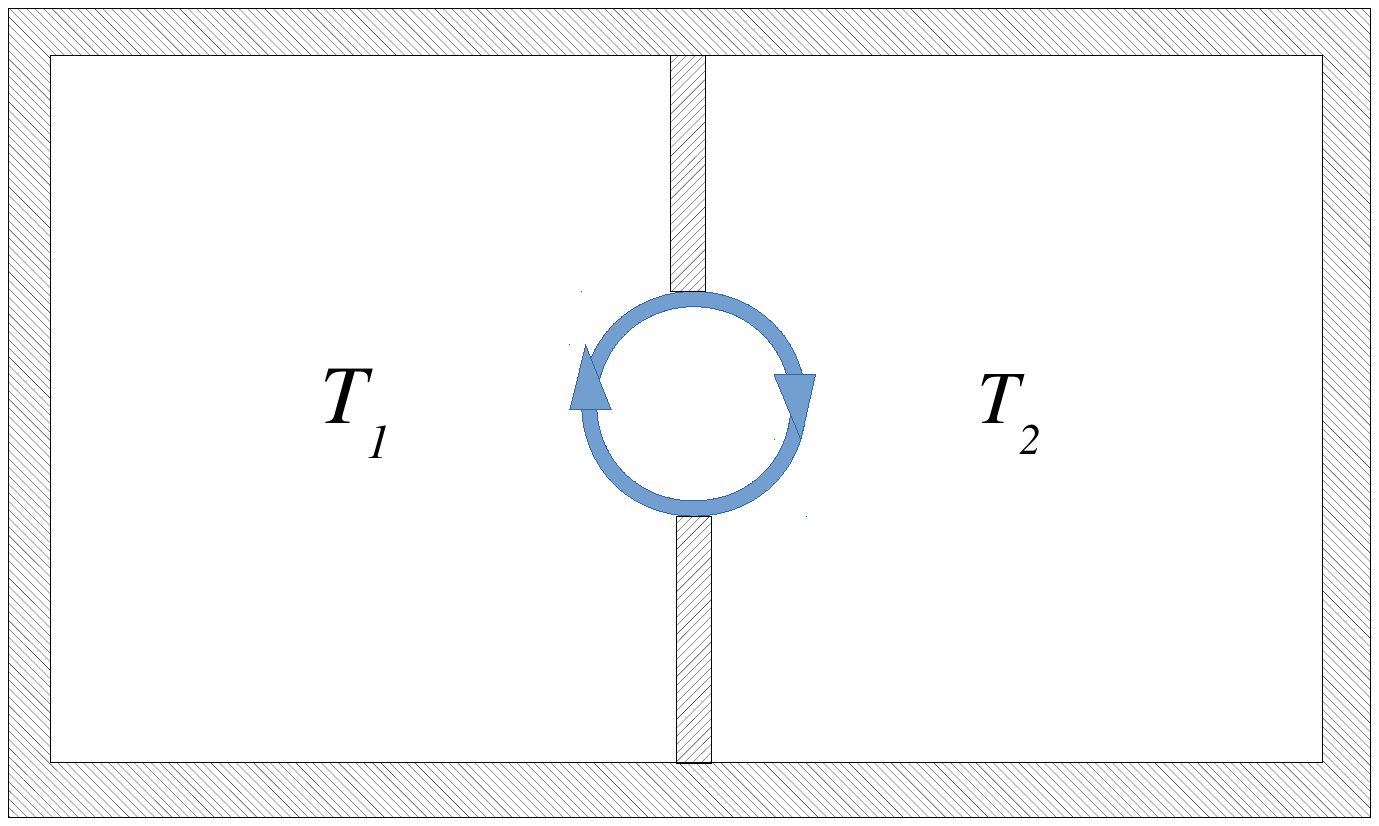}
        \caption{Two systems in contact at temperature $T_1$ and $T_2$ can be
used to extract useful work by means of an auxiliary body - depicted as circular double arrow -
making infinitesimal cyclic transformations.}
        \label{figura1}
\end{figure}

The example of the two systems offers the opportunity to show that the exergy 
defined by the equation \eqref{exeriso} is indeed the maximum available work which 
can be extracted from two finite thermal sources at different temperatures. 
To calculate the maximum work from the system in fig.~\ref{figura1} in the traditional 
textbook approach, we need to introduce an auxiliary
body which is coupled to the two systems at different temperatures (represented 
by the loop in figure~\ref{figura1}) and whose purpose is to convert some internal
energy into kinetic energy, through some coupling to an external device. The body 
is indeed {\em the} system making a Carnot cycle between the sources at different 
temperatures. In thermodynamics textbooks the sources are thermostats, while in our 
case they are finite and once some internal energy is extracted from the source at 
higher temperature, its temperature does not stay constant but it decreases,
thus we have to consider infinitesimal Carnot cycles such that the temperatures
can be considered as constant throughout them, and integrate. The work $\di L$ 
extracted in an infinitesimal cycle reads: 
$$
 \di L = - \di U_2 - \di U_1
$$
where $\di U_i$ are positive when internal energy of either source increases. 
According to the Carnot theorem, the work is maximal when the process is reversible, 
that is when the total entropy increase is zero:
$$
 \di S_{\rm tot} = \di S_1 + \di S_2 = \frac{\di U_2}{T_2} + \frac{\di U_1}{T_1} 
 = 0
$$ 
This condition allows to write the work as:
$$
 \di L = - \left(1-\frac{T_2(t)}{T_1(t)} \right) \di U_2 = - \eta \; \di U_2 
$$
where $\eta$ is the efficiency. The total work is obtained 
by integrating in time the internal energy variations between the initial state
and the final state:
$$
  L = \int_i^f \di L = - \int_i^f \di U_2 + \di U_1
$$
where the integration is constrained by $S_1+S_2={\rm constant}$. In the initial
state, the total internal energy is $U_1(T_1(0)) + U_2(T_2(0))$ whereas in the final 
state, the internal energy must be that of a thermalized system at a common temperature
with the same total entropy of the initial state, that is just $U_{\rm th}(S_1+S_2)$. 
Therefore:
$$
 L = U_1 + U_2 - U_{\rm th}(S_1+S_2)
$$
which coincides with the exergy expression \eqref{exer2}.

\section{A system in contact with a reservoir}
\label{reservoir}

\begin{figure}
       \includegraphics[scale=0.4]{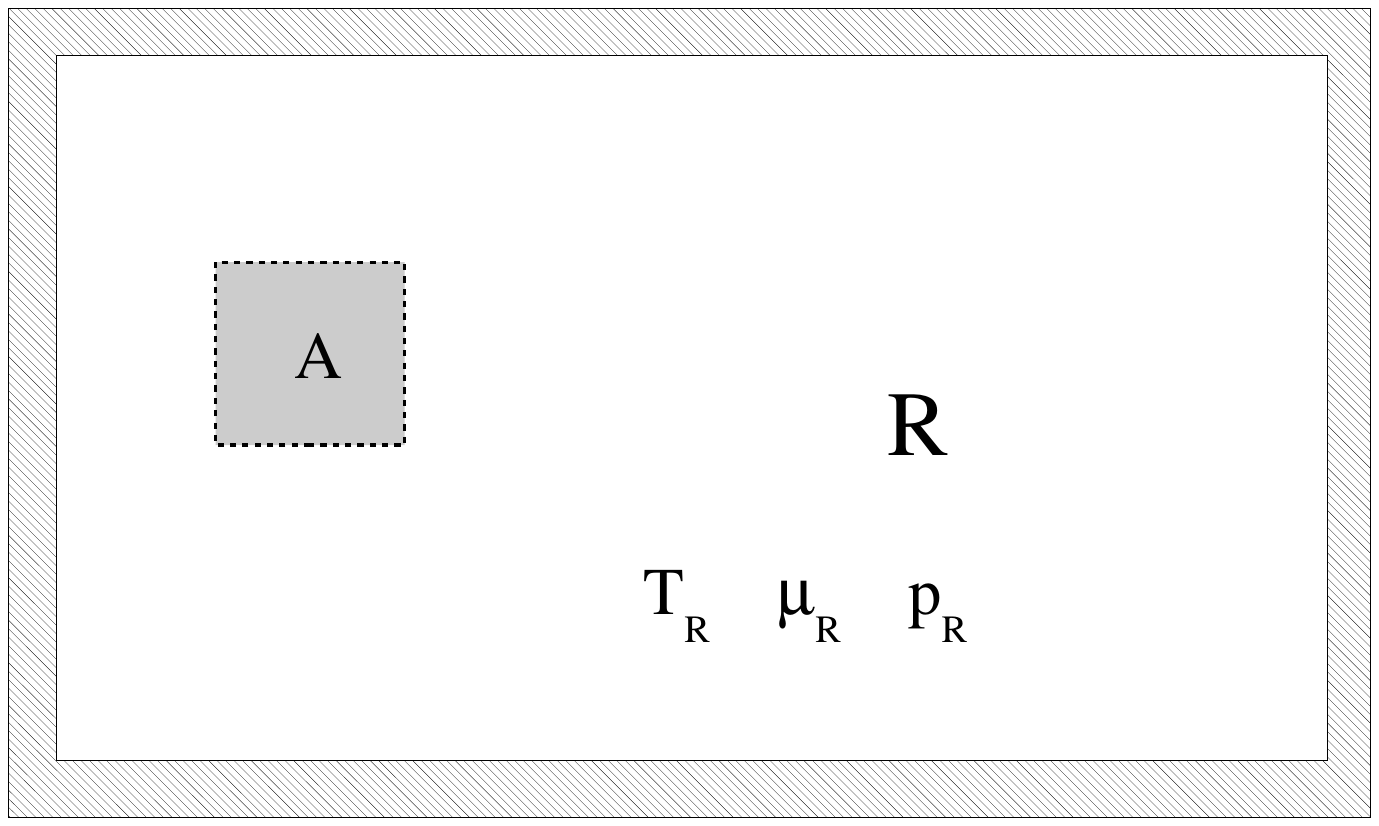}
        \caption{A small system $A$ in a thermal bath of the reservoir $R$ with
temperature $T_R$, pressure $p_R$ and chemical potentials $\mu_R$.}
        \label{figura2}
\end{figure}

Let us now consider a small system $A$ embedded in a reservoir $R$ and suppose that
they are overall isolated. The system $A$ should be viewed as a material system which
is able to move around within the environment (the Lagrangian description in the
mechanics of continuous media). The reservoir $R$ is at thermodynamic equilibrium 
with uniform and constant intensive parameters (see figure~\ref{figura2}) but the 
small system $A$ is not at thermodynamic equilibrium with $R$. The calculation of 
the exergy of $A+R$ is made it easier by the fact that the temperature $T_*(t)$ is, 
to an excellent approximation, equal to the reservoir temperature $T_R$, so that 
the equation \eqref{exergrate} is expected to become:
\be\label{deltab}
\frac{\di B}{\di t} = - T_R \frac{\di S_{\rm tot}}{\di t} 
\implies B(0) = T_R \Delta S_{\rm tot}
\ee
by using the final condition $B(\infty)=0$.

Since $A+R$ is overall isolated, the total energy $E_0$ as well as the volume
$V_A+V_R$ is constant. Suppose also that the entropy and the energy of the system 
and the reservoir are additive, a familiar case in classical thermodynamics, i.e.
there are no long-range interactions between $A$ and $R$. 
We denote with capital letters the actual thermodynamic functions of the system 
and reservoir (e.g., $S_A, V_R$), and with overlined capital letters the values 
of those functions that system and reservoir had if they were in a mutual 
thermodynamic equilibrium state with fixed total entropy $S_A+S_R$. We stress, for
the sake of clarity, that this thermodynamic equilibrium state is virtual, so to
speak, and it does not imply an actual physical equilibration process between
system and reservoir, where total entropy increases. According to the 
general definition of exergy \eqref{exergy2} for an isolated system, and taking 
into account the assumed additivity of energy, we can write the thermalized energy 
of the system {\rm and} the reservoir as:
\be\label{utherm}
 U_{\rm th}(S = S_A + S_R, V = V_A + V_R,\ldots) 
 = U_A (\overline S_A,\overline V_A,\overline N_{kA}) + 
  U_R (\overline S_R,\overline V_R,\overline N_{kR}) 
\ee
where $U$ is the equilibrium thermodynamic function internal energy and the dots 
on the left hand side stand for all other relevant constant arguments (e.g. those 
related to chemical reactions) as discussed in Section~\ref{isolated}. The $N_k$
are the absolute numbers of some chemical species, with $k=1,2,\ldots,M$. Note that:
\be\label{samesum}
  S_A + S_R = \overline S_A + \overline S_R  \qquad V_A + V_R = \overline V_A + 
  \overline V_R 
\ee
for the constancy of the volume and because the internal energy must be evaluated
at fixed entropy. 

Now, at equilibrium, the derivatives of the internal energy with respect to entropy,
volume and all the number of species are the same for system and reservoir. Besides,
they are equal to the derivatives of $U_{\rm th}$ because of the \eqref{utherm} 
and \eqref{samesum}: 
\be\label{equilib}
  \frac{\partial U_A}{\partial \overline S_A} = \frac{\partial U_R}{\partial \overline S_R}
  =  \frac{\partial U_{\rm th}}{\partial S}
\qquad \qquad 
 \frac{\partial U_A}{\partial \overline V_A} = \frac{\partial U_R}{\partial \overline V_R}
 =  \frac{\partial U_{\rm th}}{\partial V}
\qquad \qquad 
 \frac{\partial U_A}{\partial \overline N_{kA}} = \frac{\partial U_R}{\partial \overline N_{kR}}
 =  \frac{\partial U_{\rm th}}{\partial N_k}
\ee
Note that the above derivatives are calculated at the overlined equilibrium values,
which differ from the actual ones. Nevertheless, for the reservoir the difference
between the derivatives at the equilibrium value and those at the actual values
are negligible, because the reservoir is much larger than the system. Therefore
we can approximate:
\be\label{reservint}
  \frac{\partial U_R}{\partial \overline S_R} 
  \simeq \frac{\partial U_R}{\partial S_R} = T_R \qquad \qquad
  \frac{\partial U_R}{\partial \overline V_R} 
  \simeq \frac{\partial U_R}{\partial V_R} = -p_R \qquad \qquad
  \frac{\partial U_R}{\partial \overline N_{kR}} 
  \simeq \frac{\partial U_R}{\partial N_{kR}} = \mu_{kR}
\ee
where $T_R$ is the temperature of the reservoir, $p_R$ its pressure and $\mu_{kR}$
its chemical potentials, all of them supposedly constant and equal to their equilibrium
values. 

We can now calculate the exergy rate by using the definition \eqref{exergy2} and
taking into account the eq.~\eqref{reservint}:
\be\label{dbdt1}
  \frac{\di B}{\di t} = \frac{\di E_0}{\di t} - \frac{\di U_{\rm th}}{\di t} =
  0 -\frac{\partial U_{\rm th}}{\partial S} \frac{\di S}{\di t} =
  - T_R \left( \frac{\di S_A}{\di t} + \frac{\di S_R}{\di t} \right)
\ee
which is the equation \eqref{deltab}.

Since the reservoir is itself at thermodynamic equilibrium, the time derivative 
of the entropy of the reservoir can be calculated by using the known equilibrium
relations of thermodynamics:
\be\label{entrodev}
  \frac{\di S_R}{\di t} = \frac{1}{T_R} \frac{\di U_R}{\di t} + 
   \frac{p_R}{T_R} \frac{\di V_R}{\di t} - \sum_k \frac{\mu_{kR}}{T_R} 
   \frac{\di N_{kR}}{\di t} 
\ee
Taking into account the energy conservation, the volume conservation and the
general chemical species balance equations:
\begin{align}\label{balance}
 &\frac{\di E_A}{\di t} + \frac{\di U_R}{\di t} = 0 \\ \nonumber
 &\frac{\di V_A}{\di t}  + \frac{\di V_R}{\di t} = 0 \\ \nonumber
 &\frac{\di N_{kA}}{\di t} + \frac{\di N_{kR}}{\di t} = R_{kA} = \nu_k R
\end{align}
where $E_A$ is the energy of the subsystem $A$ including its kinetic collective
energy $K_A$ and its internal energy $U_A$; $R$ is the chemical reaction rate 
and $\nu_k$ the stoichiometric coefficients, negative for the reactants and positive 
for the products), the \eqref{dbdt1} can be transformed into, by using the 
\eqref{entrodev} and \eqref{balance}:
\be\label{dbdt2}
  \frac{\di B}{\di t} = \frac{\di E_A}{\di t} - T_R \frac{\di S_A}{\di t} 
  + p_R \frac{\di V_A}{\di t} - \sum_k \mu_{kR} \left( \frac{\di N_{kA}}{\di t}
  - \nu_k R \right)  
\ee
The chemical reaction term indeed vanishes because the reservoir is supposedly
at equilibrium:
$$
  \sum_k \mu_{kR} \nu_k = 0
$$
and the \eqref{dbdt2} gets simplified to:
\be\label{dbdt3}
  \frac{\di B}{\di t} = \frac{\di K_A}{\di t} + \frac{\di U_A}{\di t} - 
   T_R \frac{\di S_A}{\di t} 
  + p_R \frac{\di V_A}{\di t} - \sum_k \mu_{kR} \frac{\di N_{kA}}{\di t}  
\ee
This equation can be readily integrated over some time interval:
\be\label{deltaexe}
  \Delta B = \Delta K_A + \Delta U_A - T_R \Delta S_A + p_R \Delta V_A - \sum_k 
  \mu_{kR} \Delta N_{kA}
\ee
The exergy decreases steadily until, at the time $t=\infty$, the system achieves 
full thermodynamic equilibrium with the reservoir with the same intensive quantities 
$T_R,p_R,\mu_{kR}$. Furthermore, the kinetic
energy will vanish and the whole energy of the system will be internal, related
to the other thermodynamic variables by the equilibrium thermodynamic relation
with the intensive quantities $T_R,p_R,\mu_{kR}$: 
$$
 K(\infty) = 0 \qquad \qquad
 U_A(\infty) - T_R S_A(\infty) + p_R V_A(\infty) - \sum_k \mu_{kR} N_{kA}(\infty) = 0
$$
Therefore, from the equation \eqref{deltaexe}, we obtain the total exergy at 
the initial time $t=0$:
\be\label{exeres}
  B = K_A(0) + U_A(0) - T_R S_A(0) + p_R V_A(0) - \sum_k \mu_{kR} N_{kA}(0) 
\ee
which is just the initial exergy. The equation \eqref{exeres} is indeed a well known 
formula of the exergy of a system in contact with a reservoir and has been used 
extensively \cite{dincer}. 

We note an interesting feature of the equation \eqref{exeres}.
From its general definition for an isolated system in Section~\ref{isolated}, it 
is obvious that $B$ is positive definite, and one may wonder if this holds for 
the expression \eqref{exeres}. Indeed this does, owing to the well known thermodynamic 
stability condition \cite{landau} stating that any small deviations of the extensive 
quantities for a system in contact with a reservoir are such that:
$$
\delta K_A(0) + \delta U_A(0) - T_R \delta S_A(0) + p_R \delta V_A(0) - 
 \sum_k \mu_{kR} \delta N_{kA}(0) > 0 
$$
In other words, $B=0$ is a minimum and the right hand side of the \eqref{exeres}
is always positive.

\subsection{System in an external field}

This analysis can be extended to the case where the system $A$ (but not $R$) is 
subjected to an external field, like gravity acceleration. In this case, the 
energy of the system, defined as the sum of kinetic and internal energy \footnote
{We follow the definition in ref.~\cite{lizarraga}.} is no longer conserved because of
the injected external power $P_{\rm ext}$. The equation \eqref{dbdt1} is replaced by:
$$
 \frac{\di B}{\di t} = \frac{\di E_0}{\di t} - \frac{\di U_{\rm th}}{\di t} 
 = P_{\rm ext} - \frac{\di U_{\rm th}}{\di t} = P_{\rm ext} - T_R  
  \left( \frac{\di S_A}{\di t} + \frac{\di S_R}{\di t} \right)
$$
Likewise, the first equation \eqref{balance} is replaced by:
$$
 \frac{\di E_A}{\di t} + \frac{\di U_R}{\di t} = P_{\rm ext} 
$$
so that, eventually, the term $P_{\rm ext}$ gets cancelled and the equations 
\eqref{dbdt3} and \eqref{exeres} hold true. 

\section{Exergy of a continuous medium in contact with a reservoir}
\label{continres}

The equation \eqref{exeres} can be extended to a collection of small systems which
are all in contact with the same large reservoir. In this case, the equation \eqref{exeres}
shows that exergy can be considered as additive, because it is the sum of additive 
quantities multiplied by constant intensive quantities of the reservoir. This is 
a common assumption when dealing with small subsystems on Earth:
$$
  B = \sum_i B_i
$$
This assumption is broken when the subsystems are no longer small compared to the
reservoir or, tantamount, when the reservoir is not large enough so as to maintain 
constant and uniform thermostatic properties. 

The additivity and the extensivity of all the quantities $K, U, S, V, N$ in the 
equation \eqref{exeres} also makes it possible to define an {\em exergy density} 
throughout the system and the reservoir as:
\be\label{exerdens}
   b ({\bf x},t) = \frac{1}{2} \rho v^2 + u + p_R - T_R s - \sum_k \mu_{kR} n_k
  = \epsilon + p_R - T_R s - \sum_k \mu_{kR} n_k
\ee
where $\rho$ is the mass density, $u$ the internal energy density, $\epsilon = (1/2)
\rho v^2+u$ is the total energy density, $s$ the entropy density, $n_k$ the number
density of the species $k$. By taking into account that in the region occupied by 
the reservoir $b=0$, for the velocity field ${\bf v}$ vanishes the application of
the basic relation $u_R + p_R = T_R s_R + \sum_k \mu_{kR} n_{kR}$ at thermodynamic 
equilibrium, it can
be seen that the equation \eqref{exerdens} applies to every point in space, including
the reservoir. For a finite subsystem with volume $V$, its exergy can be obtained 
by integrating the \eqref{exerdens}, thus reproducing the equation \eqref{exeres}:
$$
 B_V = \int_V \di V \; b = \int_V \di V \; 
 \left( \frac{1}{2} \rho v^2 + u + p_R - T_R s -  \sum_k \mu_{kR} n_{kR} \right)
 = K_V + U_V + p_R V - T_R S_V - \sum_k \mu_{kR} N_{kV}
$$
It is now possible to determine a continuity equation for the exergy density by 
using the continuity equations of energy, entropy and number densities:
\begin{align}\label{continuity}
 &\frac{\partial \epsilon}{\partial t} + \nabla \cdot {\bf j}_E = \pi  \\ \nonumber
 &\frac{\partial s}{\partial t} + \nabla \cdot {\bf j}_S = \sigma 
  \\ \nonumber
 &\frac{\partial n_k}{\partial t} + \nabla \cdot {\bf j}_k = \nu_k r 
\end{align}
In the equation above, ${\bf j}$ are the fluxes; specifically, the flux of particles 
of species $k$ is ${\bf j}_k= n_k {\bf v}_k$ where ${\bf v}_k$ is the baricentric
velocity of the species $k$; $\sigma$ is the entropy production rate within the 
region $V$; $\pi$ is the power density of external fields, which is usually written
as:
$$
  \pi = \sum_k {\bf f}_k \cdot {\bf v}_k
$$
where ${\bf f}_k$ is the density of force (for intance, for the gravitational 
field at the Earth surface ${\bf f}_k = m_k n_k {\bf g}$) exerting on the species
$k$; and $r$ is the chemical reaction rate per unit volume. 
By taking the partial derivatives of both sides of \eqref{exerdens} and using the
\eqref{continuity} we get:
\begin{align*}
\frac{\partial b}{\partial t} &= \frac{\partial \epsilon}{\partial t} - T_R 
\frac{\partial s}{\partial t} - \sum_k \mu_{kR} \frac{\partial n_k}{\partial t} \\
 &= \pi - \nabla \cdot {\bf j}_E + T_R \nabla \cdot {\bf j}_S
  - T_R \sigma + \sum_k \mu_{kR} \nabla \cdot {\bf j}_k - \sum_k \mu_{kR} \nu_k r 
\end{align*}
The last term vanishes in the reservoir which is supposedly at chemical equilibrium 
so that:
$$
 \sum_k \mu_{kR} \nu_k = 0
$$
We are thus left with:
\be\label{dbdtdens}
  \frac{\partial b}{\partial t} + \nabla \cdot \left( {\bf j}_E - 
   T_R {\bf j}_S - \sum_k \mu_{kR} {\bf j}_k \right) = \pi - T_R \sigma
\ee
The above equation can now be integrated over some region $V$ to obtain the exergy
balance equation. 

The equation \eqref{dbdtdens} is the general form of exergy continuity equation 
for a multi-component continuous medium and applies to a great variety of systems.
The exergy source includes a possibly positive term $\pi$, the power supplied by 
the external field, and a negative term $T_R \sigma$ due to dissipation. The energy 
flux ${\bf j}_E$ comprises all forms of energy inward/outward fluxes, including 
electromagnetic radiation or any other sort of radiation; the same applies to the 
entropy or the chemical species flux vectors.

\subsection{Example: classical fluid}

Among all systems for which the calculation of exergy balance equation is interesting, 
there is certainly a classical multi-component fluid. We refer to ref.~\cite{degroot} 
for the derivation of the expressions quoted below. Let $\rho$ be the mass density 
and ${\bf v}$ the hydrodynamic velocity field, defined as:
$$
  \rho = \sum_k m_k n_k   \qquad \qquad {\bf v} = \frac{1}{\rho} \sum_k m_k {\bf j}_k
$$
where $k$ is the index labeling the chemical species, with $m_k$ its, $n_k$ its 
number density and ${\bf j}_k = n_k {\bf v}_k$ its flux vector.
The fluxes and the entropy production rate turn out to be \cite{degroot}:
\begin{align}\label{fluxes}
 {\bf j}_E &= \frac{1}{2} \rho v^2 {\bf v} + u {\bf v} + {\bf j}_q - {\sf T} \cdot {\bf v} 
 \\ \nonumber
 {\bf j}_S &= s {\bf v} + \frac{1}{T} {\bf j}_q - \sum_k \frac{\mu_k}{T} {\bf J}_k 
 \\ \nonumber
 \sigma &= {\bf J}_q \cdot \nabla \left(\frac{1}{T}\right) - \sum_k {\bf J}_k 
 \cdot \nabla \left(\frac{\mu_k}{T} \right) + \frac{1}{T} {\sf \Pi} : \nabla {\bf v} -
 \frac{1}{T} \sum_k \mu_k \nu_k r + \frac{1}{T} \sum_k {\bf f}_k \cdot {\bf J}_k/n_k
\end{align}
where ${\sf T} = - p {\sf I} + {\sf \Pi}$ is the Cauchy stress tensor \footnote{
The Cauchy stress tensor is defined so as to fulfill the hydrodynamic equation of 
motion $\rho \di {\bf v}/\di t = {\rm div} {\sf T} + {\bf f}$ where ${\bf f}$ is
the external field. Thereby, the non-diagonal part of the stress-energy tensor 
${\sf \Pi}$ is defined with the opposite sign with respect to ref.~\cite{degroot}.}
with the diagonal pressure component $-p {\sf I}$; ${\bf j}_q$ is the heat flux
vector and ${\bf J}_k$ are the diffusion currents:
$$
 {\bf J}_k \equiv {\bf j}_k - n_k {\bf v}
$$

We can now integrate the equation \eqref{dbdtdens} to obtain the exergy rate of change 
within the region $V$. By using the Gauss theorem:
\be\label{dbdtcont}
  \frac{\di B}{\di t}\Big|_V = \frac{\partial}{\partial t} \int_V \di V \; b 
  =- \int_{\partial V} \di S \; \hat{\bf n} \cdot 
  \left( {\bf j}_E - T_R {\bf j}_S - \sum_k \mu_{kR} {\bf j}_k \right) - 
  T_R \int \di V \; \sigma + \int \di V \; \pi 
\ee
The total flux in the boundary term becomes, by using the eqs.~\eqref{fluxes}, the
decomposition of the Cauchy stress tensor and the \eqref{exerdens}:
\begin{align}\label{exflux}
 & {\bf j}_E - T_R {\bf j}_S - \sum_k \mu_{kR} {\bf j}_k \\ \nonumber
 &= \left( \frac{1}{2} \rho v^2 + u + p - T_R s - \sum_k \mu_{kR} n_k \right) 
  {\bf v} - \sum_k \left( \mu_{kR} - \frac{T_R}{T} \mu_k \right) {\bf J}_k 
 - {\sf \Pi} \cdot {\bf v} + {\bf j}_q \left( 1 - \frac{T_R}{T} \right) \\ \nonumber
 & = b {\bf v} + (p-p_R) {\bf v} - \sum_k \left( \mu_{kR} - \frac{T_R}{T} \mu_k 
  \right) {\bf J}_k 
 - {\sf \Pi} \cdot {\bf v} + {\bf j}_q \left( 1 - \frac{T_R}{T} \right)
\end{align}
The first term in the last expression is the convective exergy flow, while the
remaining terms make up a conductive contribution:
\be\label{condflow}
 {\bf J}_B \equiv \sum_k T_R \left( \frac{\mu_{k}}{T} - \frac{\mu_{kR}}{T_R} \right) 
  {\bf J}_k + (p - p_R) {\bf v} - {\sf \Pi} \cdot {\bf v} + 
  {\bf j}_q  \left( 1 - \frac{T_R}{T} \right)
\ee
wherein a chemical, mechanical and thermal conduction terms can be recognized.
Finally, plugging the eq.~\eqref{exflux} into the \eqref{dbdtcont} and by using
the entropy flux expression in \eqref{fluxes} we get:
\be\label{dbdtcont2}
 \frac{\partial}{\partial t} \int_V \di V \; b = - \int_{\partial V} \di S \; 
 \hat{\bf n} \cdot \left( \, b {\bf v} + {\bf J}_B \right)
 - T_R \int_V \di V \sigma + \int_V \di V \; \pi  
\ee
The expression \eqref{dbdtcont2} agrees with that quoted in ref.~\cite{lizarraga}. 

\section{Exergy of an open continuous medium}
\label{contin}

The above formulae are applicable to a continuous medium, or a collection of continuous
media, which are embedded in a very large reservoir with constant thermostatic properties;
in this case, exergy is additive and it is possible to define an exergy density. 
This is especially relevant for applications to small systems on the Earth surface, 
where one would naturally assume the reservoir be the atmosphere, or the atmosphere 
along with the crust outermost shell. 

Nevertheless, if the size of the systems is not much smaller than reservoir's and 
if the thermostatic properties of the reservoir are not really constant and uniform, 
the expressions obtained in Section \ref{continres} become questionable. For instance, in 
the Earth atmosphere, temperature, pressure and, to some extent, chemical potentials are 
neither constant nor uniform. We may expect that constancy and uniformity lead to 
very good approximations when the time and space scale of the phenomenon at stake
(for instance the exergy balance of a car round trip) are much smaller than those of the 
typical variation of the atmosphere parameters, but otherwise, these variations 
may play a role.
Furthermore, the calculations in Section ~\ref{continres} do not provide a definite 
prescription for the values of the intensive parameters of the reservoir; 
this is a renowned and debated problem in exergy calculations \cite{ahrendts,gaudreau,szargref} 
(see also ref.~\cite{dincer} and references therein), sometimes called {\em 
reference environment} problem. 
 
Therefore, there are very good motivations to find out an expression of the exergy 
of an open continuous system of arbitrary size, which is not in contact with a 
reservoir. These expressions should, of course, reproduce those found in Section 
~\ref{continres} in the limit of an infinite system with constant thermostatic
parameters. 

Let then $V$ (henceforth we identify the name of the region with its volume $V$) 
be a finite region. We can define its exergy extending the equation \eqref{exergy1}, 
by replacing the constant energy with the total energy within the region $E_V(t)$ 
at the time $t$, the total entropy with the entropy of the region $V$ at the same 
time $t$, that is $S_V(t)$, and the chemical constants $Q_i$ 
with those, $Q_i(t)$ corresponding to the chemical equilibrium that would be achieved
in the region $V$ if the composition was frozen at the time $t$:
\be\label{exergyv}
  B_V(t) = E_V(t) - U_{\rm th}(S_V(t),V,Q_1(t),\ldots,Q_M(t))
\ee
The definition \eqref{exergyv} is one of the main points of this work. Even though
the exergy is not additive, the definition above makes it possible to answer the 
main questions concerning the reference environment. 
In the equation \eqref{exergyv}, the chemical ``constants" $Q_i(t)$ are, as has 
been mentioned in Section~\ref{intro}, linearly dependent on the abundances, with 
coefficients $C_{ik}$:
\be\label{qlinear2}
  Q_i(t) = \sum_{k=1}^M C_{ik} N_k(t) = \sum_{k=1}^M C_{ik} {\overline N}_k(t)
\ee
where ${\overline N}_k$ are the equilibrium abundances which differ from the actual 
ones, as discussed in Section~\ref{intro}. It should be pointed out that, whereas
the abundances depend on time and so do the $Q_i$'s, the $C_{ik}$ are still constant
because they only depend on the stoichiometric coefficients of the chemical reactions.
Furthermore, while the actual $N_k(t)$'s change in time due to chemical reactions 
within $V$ and boundary fluxes as well, the $Q_i(t)$'s and the ${\overline N}_k(t)$'s 
only change because of boundary fluxes.

Let us now take the derivative of the \eqref{exergyv}:
$$
  \frac{\di B_V}{\di t} = \frac{\di E_V}{\di t} - T_{*V}(t) \frac{\di S_V}{\di t}
- \sum_i \mu_{*iV}(t) \frac{\di Q_{i}}{\di t}
$$
where $T_{*V}(t)$ (which has been discussed in Section~\ref{isolated}) is the 
temperature and $\mu_{*iV}$ are the chemical potentials that the region $V$ would 
have if at global equilibrium with the given parameters at the time $t$. These
intensive thermodynamic parameters are defined {\em reference parameters}. By using
the \eqref{qlinear2} and the \eqref{chempotrel}, the above equation becomes:
\begin{align*}
  \frac{\di B_V}{\di t} & = \frac{\di E_V}{\di t} - T_{*V}(t) \frac{\di S_V}{\di t}
- \sum_{ik} \mu_{*iV}(t) C_{ik} \frac{\di N_k}{\di t}\\
 & = \frac{\di E_V}{\di t} - T_{*V}(t) \frac{\di S_V}{\di t}
- \sum_{k} \mu_{*kV}(t) \frac{\di N_k}{\di t}
\end{align*}
Let us now introduce the densities:
\be\label{dbvdt}
 \frac{\di B_V}{\di t} = \int_V \di V \; \left(\frac{\partial \epsilon}{\partial t} 
 - T_{*V}(t) \frac{\partial s}{\partial t} - \sum_k \mu_{*kV}(t) 
  \frac{\partial n_k}{\partial t} \right)
\ee
and plug the equations~\eqref{continuity} in the above equations to obtain the 
rate of change of the exergy $B_V$ in time by using the Gauss theorem:
\begin{align}\label{dbvdt2}
  \frac{\di B_V}{\di t} &= - \int_{\partial V} \di S \; \hat{\bf n} \cdot 
  \left( {\bf j}_E - T_{*V}(t) {\bf j}_S - \sum_k \mu_{*kV}(t) {\bf j}_k \right) - 
  T_{*V}(t) \int_V \di V \; \sigma - \sum_k \mu_{*kV} \nu_k \int_V \di V \; r + 
  \int_V \di V \; \pi \\ \nonumber
&=- \int_{\partial V} \di S \; \hat{\bf n} \cdot 
  \left( {\bf j}_E - T_{*V}(t) {\bf j}_S - \sum_k \mu_{*kV}(t) {\bf j}_k \right) - 
  T_{*V}(t) \int_V \di V \; \sigma + 
   \int_V \di V \; \pi 
\end{align}
taking into account that, by definition, the chemical potentials $\mu_{*kV}(t)$ are
those of chemical equilibrium at the time $t$ within the region $V$, so for a chemical reaction
they always fulfill $\sum_k \mu_{*kV}(t) \nu_k = 0$. The equation \eqref{dbvdt2} differs 
from the \eqref{dbdtcont} just by the replacement of $T_R$ with $T_{*V}$ and $\mu_{kR}$ 
with $\mu_{k*V}$. 

The equation \eqref{dbvdt2} expresses the rate of change of the exergy of an open 
continuous medium. In principle, by integrating it in time, with known limiting 
condition at $t \to \infty$, it is possible to calculate the exergy at the present 
time $t$. In fact, such an integration requires the knowledge of the reference 
parameters as a function of time, which are highly non-trivial to calculate. These 
$M+1$ unknowns, as we have seen, are redundant because of the $L$ chemical equilibrium 
conditions:
$$
  \sum_k \mu_{*kV} \nu_k = 0
$$
for each chemical reaction; in fact, the number of independent variables is just 
$M+1-L$. The equations which are needed to determine these independent variables 
include the total entropy constraint, that is:
\be\label{entconst}
 S(T_{*V}(t),\mu_{*1V}(t),\ldots,\mu_{*M V}(t)) = \int \di V \; 
  s(T,\mu_1,\ldots,\mu_M)
\ee
and the $M-L$ equations \eqref{qlinear2}:
$$
 Q_i = \sum_{k=1}^M C_{ik} N_k(t) = \sum_{k=1}^M C_{ik} {\overline N}_k(t)
 (T_{*V}(t),\mu_{*1V}(t),\ldots,\mu_{*M V}(t))
$$
which equate the chemical "constants" $Q_i(t)$ to their expressions as a function
of the equilibrium abundances. 
In the equation \eqref{entconst} the integrand function on the right hand side
is the entropy density as a function of the actual temperature and chemical potentials.
It is apparent that the solution of the \eqref{entconst} requires the knowledge of
the entropy density as well as the inversion of the global equilibrium entropy 
function on the left hand side with respect to intensive variables at equilibrium. 
Such a calculation may not be easy to work out.

\subsection{Example: classical fluid}

To specify the eq.~\eqref{dbvdt2} for a classical fluid, we can go through the 
same steps of the derivation in the previous section. After introducing the pressure 
$p_{*V}$ which is defined as the equilibrium pressure (at the time $t$):
$$
  p_{*V}(t) \equiv p (T_{*V}(t),\mu_{*1V}(t),\ldots,\mu_{*MV}(t))
$$
and using the equations \eqref{fluxes}, we can turn the \eqref{dbvdt2} into: 
\be\label{dbvdt3}
  \frac{\di B_V}{\di t} = - \int_{\partial V} \di S \; \hat{\bf n} \cdot \left( 
  \, b {\bf v} + {\bf J}_B \right)
 - T_{*V}(t) \int \di V \sigma + \int \di V \; \pi 
\ee
where the exergy density $b$ in the convective flow is now defined as: 
\be\label{exerdens2}
 b = \frac{1}{2} \rho v^2 + u + p_{*V} - T_{*V} s - \sum_k \mu_{*kV} n_k
\ee
and the exergy conductive flow is now:
\be\label{condflow2}
 {\bf J}_B \equiv \sum_k T_{*V}\left( \frac{\mu_{k}}{T} - \frac{\mu_{*kV}}{T_{*V}} \right) 
  {\bf J}_k + (p - p_{*V}) {\bf v} - {\sf \Pi} \cdot {\bf v} + 
  {\bf j}_q  \left( 1 - \frac{T_{*V}}{T} \right)
\ee
where both the convective and the conductive flows in the \eqref{condflow2} are 
obtained by replacing:
\be\label{repla}
  T_R \longrightarrow T_{*V} \qquad p_R \longrightarrow p_{*V}
  \qquad \mu_{kR} \longrightarrow \mu_{*kV}
\ee
in the equations \eqref{exerdens} and \eqref{condflow}.
Although the function $b$ in the definition \eqref{exerdens2} is seemingly an 
exergy density, it is in fact {\em not} a proper density because the reference parameters 
$T_{*V},p_{*V},\mu_{*kV}$ depend, in a highly non-trivial fashion, on the whole
region $V$ and are therefore not local functions of ${\bf x}$. This is at variance
with the reservoir case, where the reference parameters are permanent unchangeable constants.

By using the expression \eqref{exerdens2} in the \eqref{dbvdt}, another interesting 
equation is obtained:
$$
  \frac{\di B_V}{\di t} = \frac{\partial}{\partial t} \int_V \di V \; b 
  - \frac{\di p_{*V}}{\di t} V + \frac{\di T_{*V}}{\di t} S_V + 
   \sum_k \frac{\di \mu_{*kV}}{\di t} N_{kV}   
$$
which clearly shows that the total exergy of an open continuous medium occupying
a region $V$ is not the integral of some function over the same region. 
Nevertheless, by equating the above expression to the eq.~\eqref{dbvdt3} one obtains:
\be\label{dbvdt4}
 \frac{\partial}{\partial t} \int_V \di V \; b = 
 - \int_{\partial V} \di S \; \hat{\bf n} \cdot \left(\, b {\bf v} + {\bf J}_B \right)
 - T_{*V}(t) \int \di V \sigma + \int \di V \; \pi + \frac{\di p_{*V}}{\di t} V 
  - \frac{\di T_{*V}}{\di t} S_V - \sum_k \frac{\di \mu_{*kV}}{\di t} N_{kV}  
\ee
which has the form of a continuity equation, like in the case of a region in contact 
with a reservoir eq.~\eqref{dbdtcont2}, with additional source terms.
Indeed, the formula \eqref{dbvdt4} extends the \eqref{dbdtcont2} to the case of a 
general open continuous medium which is not in contact with a reservoir with known 
and constant reference parameters and it is one of the main results of this work. 
If $b$ is interpreted as an exergy density that the system had if the reference 
parameters $p_{*V},T_{*V},\mu_{*kV}$ were those of a reservoir, the equation \eqref{dbvdt4} 
states that the exergy varies in time because of entropy production, external power {\em and} 
because of the variation of the reference parameters. This equation is especially useful 
for the calculation of chemical exergy resources on Earth: their exergetic value 
is affected by the change of the reference parameters, including temperature, 
pressure and chemical potentials.

\section{The reference pressure and temperature of the atmosphere}
\label{atmosph}

We now present an explicit calculation of the parameters $T_*,p_*$ and $\mu_{*k}$ 
by solving the \eqref{entconst} for the Earth atmosphere, considered as an open 
continuous system. Needless to say, this is an important calculation, as the
atmosphere is usually seen as {\em the} reservoir for most exergy applications 
and it is thus crucial to have well-founded values for those parameters.
We will present here a calculation with the assumption of no chemical reactions 
between the chemical species in the atmosphere. This assumption entails a very 
accurate approximation as long as the time scale in the possible applications is much 
shorter than the composition-changing chemical reactions time. With no chemical
reactions, the atmosphere is thus in chemical equilibrium and the number of independent
thermodynamic variables is equal to the number of species in the atmosphere, plus 
of course the temperature.
By fixing the {\em relative} fractions $x_i = n_i/n$ of each gas species, the number 
of independent variables boils down to two: temperature $T$ and density $n$, or, 
tantamount, the temperature $T$ and the pressure $p$.

\begin{figure}
       \includegraphics[scale=0.4]{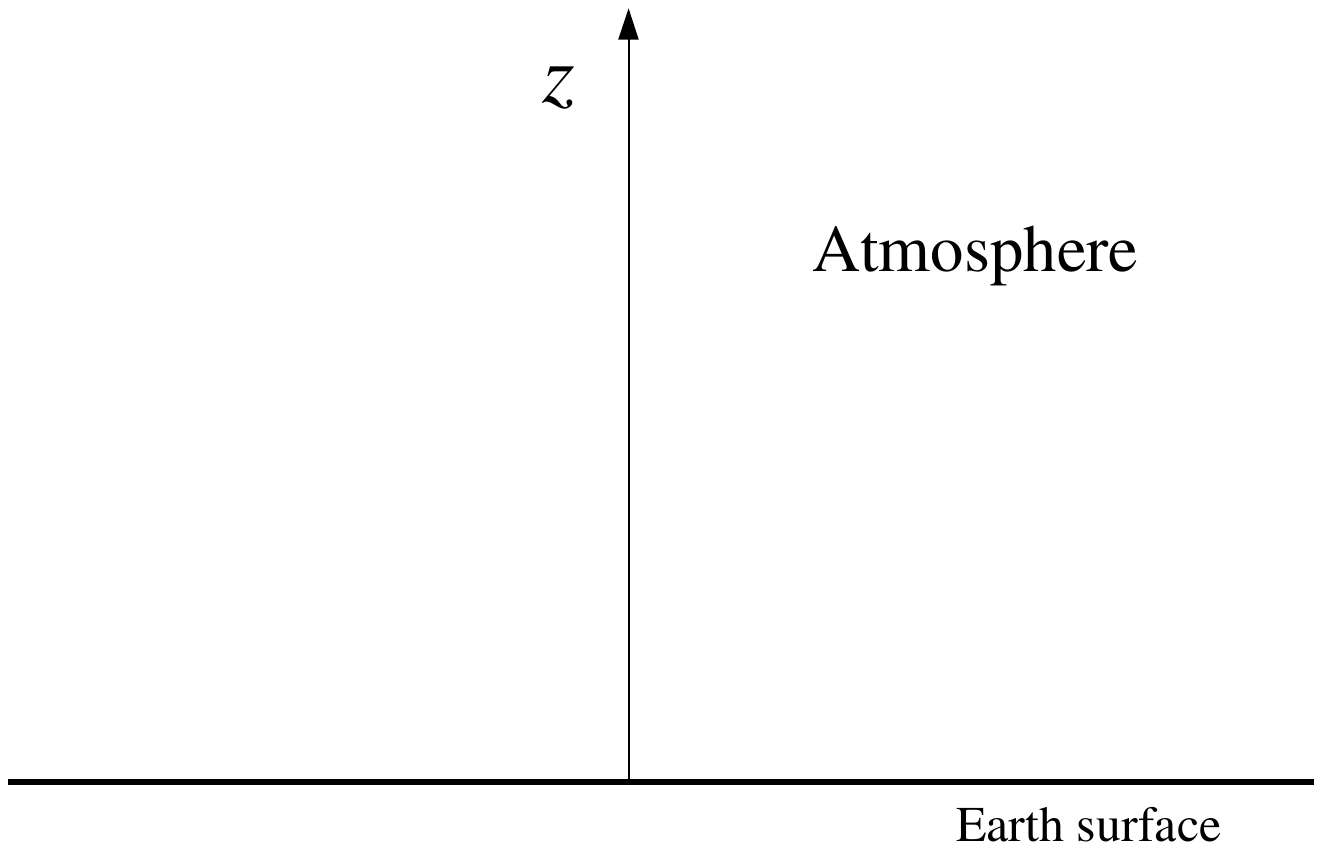}
\caption{The reference frame for the atmosphere. The coordinate $z$ is the height, 
the plane $z=0$ being the Earth surface is the lower boundary.}
        \label{figura3}
\end{figure}

The atmosphere is described as a continuous system having its lower boundary at the 
Earth surface and with no upper boundary (see figure~\ref{figura3}). Indeed, the 
gases making the atmosphere are subjected to a gravitational field and their
density decreases exponentially for increasing height, hence there is no need of 
setting an upper boundary. Even though the Earth surface is spherical, it is a very 
good approximation to describe it as a flat plane with a vertical coordinate $z$ 
like in fig.~\ref{figura3}), because the effective thickness of the atmosphere is 
much smaller than the scale of variation of the gravitational field; in other words, 
the gravity acceleration can be considered as constant throughout the atmosphere.
Furthermore, the intensive parameters are a function of the point, even at global 
equilibrium. Therefore, the equations to be solved to determine the two independent 
thermodynamic variables should be then written in principle as:
\begin{align}\label{tobesolved}
  & \int_V \di V \; s(T_*({\bf x}),p_*({\bf x})) = \int_V \di V \; s(T({\bf x}),p({\bf x}))
   \\ \nonumber
  & \int_V \di V \; n(T_*({\bf x}),p_*({\bf x})) = \int_V \di V \; n(T({\bf x}),p({\bf x}))
\end{align}
where the first corresponds to the \eqref{entconst} and the second to the constraint
of given number of molecules; the integration region $V$ corresponds to $z>0$ in
figure~\ref{figura3}.
In fact, the constraint of global equilibrium in the atmosphere demands $T_*$ to 
be uniform and the pressure $p_*$ to be a function of the vertical coordinate $z$ 
(see figure~\ref{figura3}) through:
\be\label{presseq}
 p_*({\bf x}) = p_0 \exp[- \langle m \rangle g z/T_*]   
\ee
where $p_0$ is a constant and $\langle m \rangle$ is the average mass:
$$
  \langle m \rangle \equiv \sum_k m_k \frac{n_k}{n} = \sum_k m_k x_k
$$
$m_k$ being the mass of the molecular species $k$.

The most convenient method to solve the first equation \eqref{tobesolved} is to 
make use of the so-called $TdS$ equation:
$$
  T \di S = C_p \di T - T \frac{\partial V}{\partial T}\Big|_p \di p
$$
which, in this form, just applies to a thermodynamic system with fixed chemical composition.
For point-dependent functions should use the corresponding $TdS$ equation involving 
{\em mass} densities instead:
\be\label{tds}
  T \di {\bar s} = {\bar c}_p \di T - T \frac{\partial {\bar v}}{\partial T} \di p
\ee
where the mass densities ${\bar w}$ are related to the proper densities $w$ through
the relation:
$$
  w = \rho \bar w
$$
In the eq.~\eqref{tds} $\bar c_p$ is the heat capacity per unit mass at constant
pressure (specific heat) and, obviously, $\bar v = 1/\rho$. 
We can now write down the local equation of state in terms of the usual densities:
$$
 p = n T = \sum_k n_k T  
$$
Taking into account that $\rho = \sum_k m_k n_k$ we have:
$$
  \frac{p}{\rho} = p \bar v = \frac{\sum_k n_k}{\sum_k m_k n_k} T \equiv 
  \sum_k \bar \chi_k T
$$
where $\bar \chi_k$ are the concentrations per unit mass, as it appears from the above
equation. Hence:
\be\label{derv}
  \frac{\partial \bar v}{\partial T}\Big|_p = \frac{\sum_k  \bar \chi_k}{p}
\ee
Also, from the known expression of internal energy of an ideal gas:
$$
 u = \sum_k \frac{f_k}{2} n_k T  \implies \bar u = \sum_k \frac{f_k}{2} \bar \chi_k T
$$
where $f_k$ is the number of microscopic degrees of freedom of the $k$-th molecular
species, we can readily find, by using the equation of state:
\be\label{cp}
 {\bar c}_p = \frac{\partial}{\partial T}( \bar u + p \bar v)\Big|_p = 
 \sum_k \left( \frac{f_k}{2}+1 \right) \bar \chi_k 
\ee
Therefore, by using the \eqref{derv} and the \eqref{cp}, the \eqref{tds} becomes:
\be\label{tds2}
  \di {\bar s} = \sum_k \left( \frac{f_k}{2}+1 \right) \bar \chi_k 
  \frac{\di T}{T} - \sum_k \bar \chi_k \frac{\di p}{p}
\ee
which can be integrated to give:
$$
  \bar s = \sum_k \left( \frac{f_k}{2}+1 \right) \bar \chi_k \log T - 
 \sum_k \bar \chi_k \log p  + \bar s_0
$$
taking into account that $\bar \chi_k$ are fixed. After multiplication by $\rho$, 
and using the equation of state one obtains the entropy density:
\begin{align}\label{entrodens}
  s &= \sum_k \left( \frac{f_k}{2}+1 \right) n_k \log T - n \log p
  + s_0 = \sum_k \left( \frac{f_k}{2}+1 \right) x_k n \log T - n \log p
  + s_0 \nonumber  \\
  & = \kappa n \log T - n \log p + s_0
\end{align}
where $\kappa$ is constant. The \eqref{entrodens} can be finally used in the 
equation \eqref{entconst}. The $s_0$ depends on some standard thermodynamic state, 
which can be chosen to be the exactly $T_*$ and $p_*$ - at the same point ${\bf x}$ 
where we the entropy density is calculated - as in the left hand side of the 
first equation in eq.~\eqref{tobesolved}. Therefore, this constant will 
cancel out in the first of the equations \eqref{tobesolved} and we are left with:
\be\label{firstone}
 \int_V \di V \; \frac{p_*({\bf x})}{T_*} \left( \kappa \log T_* - \log p_*({\bf x}) 
  \right) = 
 \int_V \di V \; \frac{p({\bf x})}{T({\bf x})} \left( \kappa \log T({\bf x}) - 
 \log p ({\bf x}) \right) 
\ee
to be solved along with the second of the equations~\eqref{tobesolved} which becomes, 
by using the equation of state:
\be\label{second}
 \int \di V \; \frac{p_*({\bf x})}{T_*} = \int \di V \; \frac{p({\bf x})}{T({\bf x})} 
 = N
\ee
with $N$ the total number of molecules in the atmosphere. Note that, in spite of
the appearance, these two equations are indepedent of the units chosen for the 
pressure and the temperature.

By using the eq.~\eqref{presseq}, and approximating $\di V \; = \di S \di z$ we
obtain from the \eqref{second}:
$$ 
      p_0 = \frac{N \langle m \rangle g}{A} 
$$
where $A$ is the Earth surface. The parameter $p_0$ is then independent of the 
$T_*$ and it is just the atmospheric pressure at sea level. We can also use the 
\eqref{presseq} in the equation \eqref{firstone} to obtain, after integration in $z$:
$$
 N \left( \kappa \log T_* - \log p_0 + 1 \right) = \int_V \di V \; \frac{p({\bf x})}
 {T({\bf x})} \left( \kappa \log T({\bf x}) - \log p({\bf x}) \right)
$$
whence we can determine $T_*$:
\be\label{tstar}
  T_* = \left(\frac{p_0}{\e}\right)^{1/\kappa} \exp\left[ \frac{1}{N} \int_V \di V \; 
  \frac{p({\bf x})}{T({\bf x})} \log \left( \frac{T({\bf x})}{p({\bf x})^{1/\kappa}} 
  \right) \right]
\ee
This is the final formula for the reference temperature of the atmosphere. To calculate
it, we have to integrate the temperature and pressure over the whole atmosphere
volume, i.e. $z>0$ in figure~\ref{figura3}. Since $p({\bf x})/T({\bf x}) = n({\bf x})$ 
one can see that, expectedly, the region near $z=0$ with higher density weigh more 
in the integral in the eq.~\eqref{tstar} than those at lower density located at 
high $z$. The constant $\kappa$:
$$
\kappa = \sum_k \left( \frac{f_k}{2}+1 \right) x_k 
$$
can be readily calculated from the composition of the atmosphere and its approximate
value is $\simeq 3.55$. Even if not manifestly, it can be checked that the formula 
\eqref{tstar} turns out to be independent of the chosen temperature and pressure units. 

A more suggestive form of the \eqref{tstar} can be found by reintroducing the 
density $n$:
$$
  T_* = \left(\frac{p_0}{\e}\right)^{1/\kappa} \exp\left[ \frac{1}{N} 
 \int_V \di V \; n({\bf x}) \log \left( \frac{T({\bf x})}{p({\bf x})^{1/\kappa}} \right) \right] 
  = \left(\frac{p_0}{\e}\right)^{1/\kappa} \exp\left[\int_V \di V \; 
  \log \left( \frac{T({\bf x})}{p({\bf x})^{1/\kappa}} \right)^{n({\bf x})/N} \right] 
$$
We can now approximate the integration with a discrete sum over sufficiently small
cells and we have:
$$
  T_* \simeq \left(\frac{p_0}{\e}\right)^{1/\kappa} 
 \exp\left[ \sum_{i} \Delta V_i \log \left(\frac{T_i}{p_i^{1/\kappa}}\right)^{n_i/N} \right] 
 = \left(\frac{p_0}{\e}\right)^{1/\kappa} 
 \exp\left[ \sum_{i} \log \left(\frac{T_i}{p_i^{1/\kappa}}\right)^{n_i \Delta V_i/N} \right]
$$
whence
\be\label{tstar2}
 T_* = \left(\frac{p_0}{\e}\right)^{1/\kappa} \prod_i 
  \left(\frac{T_i}{p_i^{1/\kappa}}\right)^{N_i/N}
\ee
where $N_i = n_i \Delta  V$ is the number of molecules in the $i$-th cell. 
The equation \eqref{tstar2} tells us that, surprisingly, the exergy reference temperature of 
the atmosphere is to be calculated with a modified weighted {\em geometric mean}, 
and not with a weighted linear mean as one would have naively guessed. This result
is in agreement with calculations of the reference temperature in ref.~\cite{grubb}
for a collections of ideal gas cells at different temperatures and constant densities.

The numerical computation of the right hand side of the eq.~\eqref{tstar2} requires 
the analysis of the recorded temperatures and pressures at different locations 
and altitudes. We can take the logarithm of both sides of \eqref{tstar2} and write:
\be\label{logstar}
  \log T_* = - \frac{1}{\kappa} + \sum_i \frac{N_i}{N} \log T_i -   
 \frac{1}{\kappa} \sum_i \frac{N_i}{N} \log \left(\frac{p_i}{p_0} \right)
\ee
Let now be $\bar T$ the average temperature calculated with a weighted arithmetic mean:
$$
  \bar T  = \sum_i \frac{N_i}{N} T_i
$$
with: 
$$
 T_i = \bar T + \delta T_i
$$
which is just a definition of $\delta T_i$. Similarly, we can define a pressure
fluctuation $\delta p_i$ with respect to a most probable value as expressed by
the equation \eqref{presseq}:
$$
 p_i = p_{*i} + \delta p_i = p_0 \exp[-\langle m \rangle g z/T_*] + \delta p_i
$$
By plugging the above definitions in the equation \eqref{logstar} we obtain:
$$
 \log T_* = \sum_i \frac{N_i}{N} \log (\bar T + \delta T_i) -   
 \frac{1}{\kappa} \left[ \sum_i \frac{N_i}{N} 
 \left(-\frac{\langle m \rangle g z}{T_*}\right)     
 +  \log \left(1 + \frac{\delta p_i}{p_{*i}} \right) +1 \right]
$$
It can now be checked, by turning the sum into an integration and using the equation
of state, as well as the equation \eqref{presseq} and the value of $p_0$, that 
the first term within the square brackets on the right hand side yields a $-1$.
We are then left with:  
$$
 \log T_* = \sum_i \frac{N_i}{N} \log (\bar T + \delta T_i) -   
 \frac{1}{\kappa} \left[ \sum_i \frac{N_i}{N} \log \left(1 + \frac{\delta p_i}{p_{*i}} 
 \right) \right]
$$
The $\delta T_i$ and $\delta p_i$ are supposedly small compared to the central
values, so we can expand to second order in the $\delta T_i,\delta p_i$:
$$
\log T_* \simeq \log \bar T - 
 \frac{1}{2} \sum_i \frac{N_i}{N} \left(\frac{\delta T_i}{\bar T}\right)^2
+ \frac{1}{2\kappa} \left[ \sum_i \frac{N_i}{N} \left(\frac{\delta p_i}{p_{*_i}}\right)^2 
\right]
$$
From this equation we can obtain an approximation of the reference temperature as:
$$
  T_* \simeq \bar T \exp\left[-\frac{1}{2} \sum_i \frac{N_i}{N} 
  \left( \frac{\delta T_i}{\bar T} \right)^2 + \frac{1}{2 \kappa} \frac{N_i}{N} 
  \left(\frac{\delta p_i}{p_{*_i}}\right)^2 \right]
 \simeq \bar T -\frac{\bar T}{2} \sum_i \frac{N_i}{N} 
  \left( \frac{\delta T_i}{\bar T} \right)^2 + \frac{\bar T}{2 \kappa} \frac{N_i}{N} 
  \left(\frac{\delta p_i}{p_{*_i}}\right)^2
$$
Hence, after defining the temperature and pressure variance as:
$$
 \sigma^2_T \equiv  \sum_i \frac{N_i}{N} \left( \frac{\delta T_i}{\bar T} \right)^2
 \qquad \qquad
 \sigma^2_p \equiv  \sum_i \frac{N_i}{N} \left(\frac{\delta p_i}{p_{*_i}}\right)^2
$$
we have:
\be\label{tstarfinal}
  T_* \simeq \bar T \left( 1 - \frac{1}{2} \sigma^2_T + \frac{\bar T}{2\kappa} \sigma^2_p
   \right)
\ee
The deviation of the reference temperature from the weighted aritmetic mean over
atmospheric cells is therefore small, as $\sigma_T$ and $\sigma_p$ are expectedly
much smaller than 1.

\section{Summary and conclusions}

We have reviewed the concept of exergy from the point of view of non-equilibrium
thermodynamics and provided a general compact definition:
$$
  B_V(t) = E_V(t) - U_{\rm th}(S_V(t),V,Q_1(t),\ldots,Q_M(t))
$$
(see Section~\ref{contin}) as the difference between the energy and the thermalized
energy within a region $V$. This definition applies to an arbitrary open continuous 
system, not necessarily in contact with an equilibrium 
reservoir or an environment with fixed intensive thermodynamic parameters. 
Such definition leads to general continuity equations \eqref{dbvdt2},\eqref{dbvdt4} 
which extend known expressions \cite{lizarraga} of continuous media in contact with
a reservoir. Besides, the general definition makes it possible to {\em calculate} the 
reference thermodynamic parameters for a system which is large enough to be considered
as an environment in many applications. As an example, we have derived the equations 
to calculate the reference parameters of the Earth atmosphere in a general non-equilibrium 
state for the pressure \eqref{presseq} and the temperature \eqref{tstar2}. 

\section*{Acknowledgments}

I am grateful to Dr. A. Giuntoli for encouragement and fruitful discussions.




\begin{thebibliography}{99}
\section*{References}

\bibitem{rant} 
Z. Rant, {\it Exergie, ein neues Wort fur "Technische Arbeitsfahigkeit" 
(Exergy, a new word for "technical available work")}. Forschung Auf dem Gebiete 
des Ingenieurwesens {\bf  22}, 36–37 (1956).

\bibitem{perot} 
 P. Perrot, {\it A to Z of Thermodynamics}, Oxford University Press (1998).

\bibitem{landau} 
 L. Landau, E. Lifshitz, {\it Statistical Physics}, Pergamon press (1980). 

\bibitem{szargut} J. Szargut, "Exergy method. Technical and ecological
applications", WIT press (2005).

\bibitem{dincer}
I. Dincer, M. Rosen "Exergy", Springer (2012).

\bibitem{lizarraga}
J. M. Sala-Lizarraga, A. Picallo-Perez, "Exergy analysis and Thermoeconomics
of buildings", Elsevier science (2019).

\bibitem{jorgensen}
S. E. Jorgensen, "Eco-exergy as sustainability", WIT press (2006).

\bibitem{stanek}
W. Stanek {\it et al.}, "Thermo-ecology: exergy as a measure of sustainability",
Elservier science (2019).

\bibitem{grubb} 
 R. W. Grubbstr\"om, Int. J. of Energy Optimization and Engineering {\bf 1}
 (2012) 1.

\bibitem{schlogl}
 F. Schl\"ogl, {\it Probability and Heat}, Springer (1989).

\bibitem{degroot}
 S. R. De Groot, P. Mazur, {\it Non-equilibrium thermodynamics}, North Holland (1969).

\bibitem{zemanski}
 M. W. Zemansky, R. Dittman, {\it Heat and Thermodynamics: An Intermediate Textbook},
  New York McGraw-Hill (1997).

\bibitem{ahrendts}
 J. Ahrendts, Energy {\bf 5} (1980) 667.

\bibitem{gaudreau}
 K.~Gaudreau, R. A. Fraser, S. Murphy, Energies {\bf 5} (2012) 2197-2213.

\bibitem{szargref}
J. Szargut, A. Valero, W. Stanek, A. Valero 
{\it Towards an international legal reference environment}, Proceedings of ECOS, 
2005, 409-420.


\end{thebibliography}
\end{document}